\documentclass[conference,a4paper]{IEEEtran}

\ifCLASSINFOpdf
\else
\fi

\usepackage{graphicx}
\usepackage{mathtools}
\usepackage{amsmath}
\usepackage{amsfonts}
\usepackage{enumerate}
\usepackage{dsfont}
\usepackage{amssymb}
\usepackage{bbold}

\newtheorem{theorem}{Theorem}[section]
\newtheorem{lemma}[theorem]{Lemma}
\newtheorem{proposition}[theorem]{Proposition}

\newenvironment{example}[1][Example]{\begin{trivlist}
\item[\hskip \labelsep {\it #1}]}{\end{trivlist}}

\newcommand{\qed}{\nobreak \ifvmode \relax \else
      \ifdim\lastskip<1.5em \hskip-\lastskip
      \hskip1.5em plus0em minus0.5em \fi \nobreak
      \vrule height0.75em width0.5em depth0.25em\fi}

\usepackage[latin1]{inputenc}
\usepackage{tikz}
\usepackage{verbatim}
\usetikzlibrary{patterns}

\newif\iffinal 

\pgfrealjobname{generalFG-ISIT-short-corrected} 

\iffinal
  
\else
  
\fi

\begin{document}
%
\title{Analysis of Coupled Scalar Systems by Displacement Convexity}

\author{\IEEEauthorblockN{Rafah El-Khatib\IEEEauthorrefmark{1},
Nicolas Macris\IEEEauthorrefmark{1},
Tom Richardson\IEEEauthorrefmark{2} and 
Ruediger Urbanke\IEEEauthorrefmark{1}}
\IEEEauthorblockA{\IEEEauthorrefmark{1}School of Computer and Communication Sciences\\
EPFL, Lausanne, Switzerland\\ Emails: \{rafah.el-khatib,nicolas.macris,ruediger.urbanke\}@epfl.ch}
\IEEEauthorblockA{\IEEEauthorrefmark{2} Qualcomm, USA\\
Email: tomr@qti.qualcomm.com}}

\maketitle

\begin{abstract}
Potential functionals have been introduced recently as an important tool for the analysis of coupled scalar systems (e.g. density evolution equations). 
In this contribution we investigate interesting properties of this potential. Using the tool of 
displacement convexity we show that, under mild assumptions on the system, the potential functional is \emph{displacement convex}. 
Furthermore, we give the conditions on the system such that the potential is \emph{strictly} displacement convex in which case the minimizer is unique.
\end{abstract} \IEEEpeerreviewmaketitle

\section{Introduction}
Spatially coupled systems have been used recently in various
frameworks, such as compressive sensing, statistical physics, coding,
and random constraint satisfaction problems (see \cite{KRU12} and
the references therein for a review of the literature). They have been shown to exhibit excellent
performance, often optimal, under low complexity message passing
algorithms. For example, spatially coupled codes achieve capacity
under such algorithms \cite{KRU12}.

The performance of these systems is assessed
by the solutions of (coupled) Density Evolution (DE) update equations. In 
general, these equations 
can be viewed as the stationary point equations of a functional 
that is typically called ``the 
potential''. It has already been recognized that this variational 
formulation is a powerful tool to analyze DE updates under 
suitable initial conditions \cite{Pfister} .
There are various possible formulations of this 
potential functional \cite{KRU12}, \cite{Pfister}, \cite{hassani2010itw}, \cite{PfisterMacris}. In this paper 
we will use the representation of \cite{KRU12} for scalar systems. 

In a previous contribution \cite{ElKhatib}, \cite{ElKhatib2} we showed that the
potential (in the form \cite{Pfister}) associated to a spatially
coupled Low-Density Parity-Check (LDPC) code whose single system
is the $(\ell,r)$-regular Gallager ensemble, with transmission over
the BEC($\epsilon$), has a convex structure called ``displacement
convexity''. This structure is well known in the theory of optimal
transport \cite{Villani}. In fact the potential we consider in
\cite{ElKhatib}, \cite{ElKhatib2} is \emph{not} convex in the usual sense but it
\emph{is} in the sense of displacement convexity. This in itself
is an interesting property.  When displacement convexity is strict
one deduces that the minimum of the potential is unique --- assuming
it exists --- and thus so is the solution of the DE equation.

The main purpose of the present note is to prove that a general class
of scalar systems also exhibits the property of displacement convexity,
and even strict displacement convexity under rather mild assumptions.
For this purpose we will use the potential in the representation of \cite{KRU12}
which allows to obtain more transparent, general, and simpler proofs.

This manuscript is organized as follows. Section~\ref{sectionSetUp}
introduces the model and the variational formulation.  In
Section~\ref{sectionRearr} we prove rearrangement inequalities that
allow us to naturally consider the potential as a functional
of cumulative distribution functions.  The potential is
shown to be displacement convex in Section~\ref{sectionDC}.  Strict
displacement convexity
and unicity of the minimizer are 
proved
in Section~\ref{sectionUnicity}.

\section{Set Up and Variational Formulation} \label{sectionSetUp}

The natural setting for displacement convexity is the continuum
case which can be thought of as an approximation of the discrete
system in the regime of large spatial length and window size. The
continuum limit has already been introduced in the literature as a
convenient means to analyze the behavior of the original discrete
model \cite{KRU12}, \cite{hassani2010itw}, \cite{HassaniMacrisUrbanke}, \cite{donoho2012information}.

Consider a spatially coupled system with an averaging window
$w:\mathds{R}\to \mathds{R}$ which is always assumed to be
bounded, non negative, even, integrable and normalized such that
$\int_{\mathds{R}} dx\, w(x) =1$ (as we will see, sometimes further
assumptions will be necessary depending on the statements).  We
denote by  $\otimes$ the standard convolution on $\mathds{R}$ and
express the ``fixed point DE equations of a scalar continuous system"
as follows:
\begin{align}
g(x)&=h_g((f\otimes w)(x)),\label{eqnUpdateEqnsG}\\
f(x)&=h_f((g\otimes w)(x)),\label{eqnUpdateEqnsF}
\end{align}
where $x\in \mathds{R}$ is the spatial position.  We will often use
the shorthand notation $f^w=f \otimes w$ and $g^w=g \otimes w$;
further, we will often refer to the functions $f$, $g$ as {\it
profiles} and to $h_f$, $h_g$ as {\it update functions}.

We will also adopt a convenient normalization for all these functions.
As will become clear in the example below it is always
possible to adopt this normalization in specific applications.
First, we assume that the profiles are bounded. Specifically, $f,
g : \mathds{R} \to [0,1]$.  Next, we assume that the update functions
$h_f$ and $h_g$ are non-decreasing bounded functions $h_{f,g}:[0,1]
\to [0,1]$ normalized such that $h_f(0)=h_g(0) =0$ and $h_f(1)=h_g(1)=1$.
We will think of them as EXIT-like curves $(u,h_f(u))$ and $(h_g(v),v)$
for $u,v\in [0,1]$ (see the generic plot). 

Consider the signed 
area between the two curves, namely
\begin{equation}\label{eqnSignedArea}
 A(h_f, h_g; u) = \int_0^u \mathrm{d}u^\prime \, (h_g^{-1}(u^\prime) - h_f(u^\prime)).
\end{equation}
This is a functional of $h_f$, $h_g$ and a function of $u\in [0,1]$.
We consider the case where $A(h_f, h_g;u) > 0$ for all $u\in ]0,1[$ and $A(h_f, h_g;1) = 0$. 
This is equivalent to the \emph{strictly positive gap condition} of \cite{KRU12}.
In \cite{KRU12} the condition was stated in terms of the function of $u$ and $v$
\[
\phi(h_f,h_g;u,v) =  \int_0^u \text{d}u'\,h_g^{-1}(u') +  \int_0^v \text{d}v' \, h_f^{-1}(v') \, - uv\,.
\]
the condition being that $\phi$ is positive for $(u,v)\in [0,1]^2$ except at $(0,0)$ and
$(1,1)$ where it takes the value $0.$
$A(h_f, h_g;u)$ is obtained by minimizing \( \phi(h_f,h_g;u,v) \) over $v.$

\begin{example}
Take a
spatially coupled LDPC code whose 
single system is the $(\ell,r)$-regular Gallager ensemble, 
with transmission over the BEC($\epsilon$).
Let $u$ (resp. $v$) be the erasure probabilities emitted by check (resp.  variable) nodes. Let the functions $\tilde h_f$ 
(resp. $\tilde h_g$)
give the usual erasure probabilities emitted on variable (resp. check) node sides. Explicitly,
 $\tilde h_f(u)=\epsilon u^{\ell-1}$ and $\tilde h_g(v)=1-(1-v)^{r-1}$, and the usual DE equations are $v=\tilde h_f(u)$, $u=\tilde h_g(v)$. We consider the special case 
 $\epsilon=\epsilon_{\text{\tiny MAP}}$. Let $(u_{\text{\tiny MAP}}, v_{\text{\tiny MAP}})$ be the unique non-trivial fixed point 
 such that $A(\tilde h_f, \tilde h_g; u_{\text{\tiny MAP}}) =0$. The {\it normalized} functions $h_f$ and $h_g$ are defined as 
 $h_f(u)= \tilde h_f( u_{\text{\tiny MAP}} u)/ v_{\text{\tiny MAP}}$ and $h_g(v)= \tilde h_g( v_{\text{\tiny MAP}} v)/u_{\text{\tiny MAP}}$.
 The coupled DE fixed point equations are 
\begin{align*}
u_{\text{\tiny MAP}}g(x)=1-\big(1-f^{w}(x)v_{\text{\tiny MAP}}\big)^{r-1},\quad f(x)=\big(g^{w}(x) \big)^{\ell-1}.
\end{align*}
\end{example}
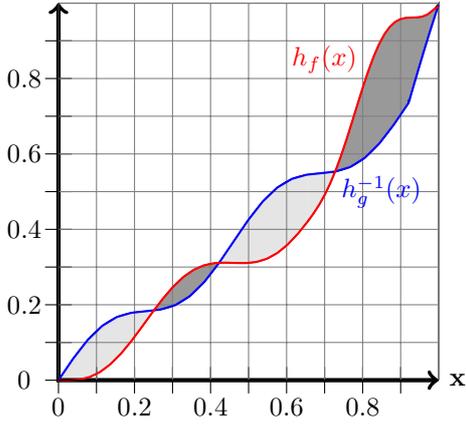
\begin{figure}
\centering
    \beginpgfgraphicnamed{plotEXITcurves-external}%
    \begin{tikzpicture}[scale=5]

\draw [fill=gray,fill opacity=0.2] plot [smooth,samples=100,domain=0:0.19/0.76](\x, {(sin(18*0.76*\x r)+20*0.76*\x)/19.08}) -- plot [smooth,samples=100,domain=0.19/0.76:0] (\x, {(-sin(0.76*18*\x r)+17*0.76*\x)/19.08});

\draw [fill=gray,fill opacity=0.8] plot [smooth,samples=100,domain=0.19/0.76:0.32/0.76](\x, {(sin(18*0.76*\x r)+20*0.76*\x)/19.08}) -- plot [smooth,samples=100,domain=0.32/0.76:0.19/0.76] (\x, {(-sin(0.76*18*\x r)+17*0.76*\x)/19.08});

\draw [fill=gray,fill opacity=0.2] plot [smooth,samples=100,domain=0.32/0.76:0.5/0.76](\x, {(sin(18*0.76*\x r)+20*0.76*\x)/19.08}) -- plot [smooth,samples=100,domain=0.5/0.76:0.32/0.76]  (\x, {(-sin(0.76*18*\x r)+17*0.76*\x)/19.08});
\draw [fill=gray,fill opacity=0.2] plot [smooth,samples=100,domain=0.5/0.76:0.55/0.76](\x, {(sin(18*0.76*\x r)+20*0.76*\x)/19.08}) -- plot [smooth,samples=100,domain=0.55/0.76:0.5/0.76] (\x, {(-sin(0.76*18*\x r)+17*0.76*\x)/19.08+(31*(0.76*\x-0.5)-sin(30*(0.76*\x-0.5) r))/19.08});

\draw [fill=gray,fill opacity=0.8] plot [smooth,samples=100,domain=0.55/0.76:0.7/0.76](\x, {(sin(18*0.76*\x r)+20*0.76*\x)/19.08}) -- plot [smooth,samples=100,domain=0.7/0.76:0.55/0.76] (\x, {(-sin(0.76*18*\x r)+17*0.76*\x)/19.08+(31*(0.76*\x-0.5)-sin(30*(0.76*\x-0.5) r))/19.08});
\draw [fill=gray,fill opacity=0.8] plot [smooth,samples=100,domain=0.7/0.76:1](\x, {(sin(18*0.76*\x r)+20*0.76*\x+39*(0.76*\x-0.7)+sin(10*(0.76*\x-0.7) r))/19.08}) -- plot [smooth,samples=100,domain=1:0.7/0.76] (\x, {(-sin(0.76*18*\x r)+17*0.76*\x)/19.08+(31*(0.76*\x-0.5)-sin(30*(0.76*\x-0.5) r))/19.08});

\foreach \x in {0,0.1,...,1}
        \draw (\x,1pt) -- (\x,-1pt); 
\foreach \y in {0,0.1,...,1}
        \draw (1pt,\y) -- (-1pt,\y);

\node (x0) at (0,-0.07) {$0$};
\node (x2) at (0.2,-0.07) {$0.2$};
\node (x4) at (0.4,-0.07) {$0.4$};
\node (x6) at (0.6,-0.07) {$0.6$};
\node (x8) at (0.8,-0.07) {$0.8$};

\node (y0) at (-0.09,0) {$0$};
\node (y2) at (-0.09,0.2) {$0.2$};
\node (y4) at (-0.09,0.4) {$0.4$};
\node (y6) at (-0.09,0.6) {$0.6$};
\node (y8) at (-0.09,0.8) {$0.8$};

\node (x) at (1.05,0) {$\mathbf{x}$};

\draw[ultra thick, <->] (0,1) -- (0,0) -- (1,0);
\draw[] (1,0) -- (1,1) -- (0,1);

\draw[gray!20!white, ultra thick] (0.5/0.76,0.428) -- (0.5/0.76,0.5447);
\draw[gray!80!white, ultra thick] (0.7/0.76,0.744) -- (0.7/0.76,0.9615);
\draw[gray!80!white, ultra thick] (0.927,0.91) -- (0.927,0.9615);

\draw[step=0.1,gray,thin] (0,0) grid (1,1);

\draw[blue,thick, domain=0:0.7/0.76] plot (\x, {(sin(18*0.76*\x r)+20*0.76*\x)/19.08});
\draw[blue,thick, domain=0.7/0.76:1] plot (\x, {(sin(18*0.76*\x r)+20*0.76*\x+39*(0.76*\x-0.7)+sin(10*(0.76*\x-0.7) r))/19.08});

\draw[red,thick, domain=0:0.5/0.76] plot (\x, {(-sin(0.76*18*\x r)+17*0.76*\x)/19.08});
\draw[red,thick, domain=0.5/0.76:1] plot (\x, {(-sin(0.76*18*\x r)+17*0.76*\x)/19.08+(31*(0.76*\x-0.5)-sin(30*(0.76*\x-0.5) r))/19.08});

\node (hf) [red, scale=1] at (0.7,0.85) {$h_f(x)$};
\node (hgInv) [blue,scale=1] at (0.85,0.5) {$h_g^{-1}(x)$};

\draw (0,0) -- coordinate (x axis mid) (1,0);
\draw (0,0) -- coordinate (y axis mid) (0,1);

\end{tikzpicture}%
    \endpgfgraphicnamed%
  
\caption{\label{plot} An example of the systems we consider. The EXIT-like curves are $h_f$ (in red) and $h_g^{-1}$ (in blue). The signed area $A(h_f,h_g;1)$ from \eqref{eqnSignedArea} is the sum of the light gray areas (positively signed) and the dark gray areas (negatively signed), and it is equal to $0$.}
\end{figure}

It is not difficult to check that the DE equations \eqref{eqnUpdateEqnsG}-\eqref{eqnUpdateEqnsF} are 
stationary point equations of a {\it potential functional of both profiles} $f$ and $g$,
\begin{align}\label{eqnPotBothFts}
\mathcal{W}(f,g)=\int\limits_{\mathds{R}}\mathrm{d}x\Big\{I_{f,g}(x)-f(x)g^w(x) \Big\}
\end{align}
where 
\begin{equation*}
I_{f,g}(x)=\int_0^{g(x)}\mathrm{d}u\, h_g^{-1}(u)+\int_0^{f(x)}\mathrm{d}v\, h_f^{-1}(v).
\end{equation*}
In order for the integral over $\mathds{R}$ in \eqref{eqnPotBothFts} to exist we have to impose 
some conditions on the profiles $f$ and $g$. These will thus be taken in the spaces (here $\epsilon >0$)
\begin{align}\label{eqnFGlimits}
\mathcal{S}_f=\{ f:\mathds{R}\rightarrow [0,1]&,\\
\;\lim\limits_{x\rightarrow -\infty}x^{1+\epsilon}f(x)&=0,\;\, \lim\limits_{x\rightarrow +\infty}x^{1+\epsilon}(f(x)-1)=0\},\nonumber\\
\mathcal{S}_g=\{g:\mathds{R}\rightarrow [0,1]&,\label{eqnFGlimitsB}\\
\;\lim\limits_{x\rightarrow -\infty}x^{1+\epsilon}g(x)&=0,\;\, \lim\limits_{x\rightarrow +\infty}x^{1+\epsilon}(g(x)-1)=0\}.\nonumber
\end{align}
%
The fixed point profile solutions of the DE equations \eqref{eqnUpdateEqnsG}, \eqref{eqnUpdateEqnsF} can be seen as the left side of the (symmetric) decoding waves.
In this paper we assume that these solutions belong to the spaces $\mathcal{S}_f$ and $\mathcal{S}_g$. 
This is achieved under some mild conditions on the slopes of the update functions $h_f$ and $h_g$ at the corner points, and if $w$ 
decays fast enough.
In particular this is true for the example of the BEC$(\ell,r)$ with a finitely supported $w$,  
where the limiting values are approached at least exponentially fast.


\section{Rearrangements}\label{sectionRearr}

Displacement convexity is defined on  a space 
 of probability measures. For measures on the real line 
 it is most convenient to view displacement 
 convexity on a space of cumulative distribution functions (cdf's).
In this section we use the tool of \emph{increasing rearrangements} to show that such rearrangements of $f$ and $g$ can only
decrease the potential. 

We first give a brief introduction to the notion of non-decreasing rearrangement, 
see \cite{Hardy}. Consider a profile $p:\mathds{R}\rightarrow[0, 1]$ 
such that $\lim\limits_{x\rightarrow -\infty}p(x)=0$ and $\lim\limits_{x\rightarrow +\infty}p(x)=1.$
Then, the {\it increasing rearrangement\footnote{Note that an increasing rearrangement is not necessarily strictly increasing.}} of $p$ is the non-decreasing function
$\bar{p}$ so that this function has the same limits and the mass of each level set is preserved. More formally, let us represent
 $p$ 
 in layer cake form as $ p(x) =\int_{0}^{p(x)}\mathrm{d}t = \int_{0}^{+\infty}\mathrm{d}t\, \chi_t(x)$, where $\chi_t(x)$ is the indicator function of the level
set $E_t=\{x\,\vert\, p(x) > t\}$.  For each value $t\in[0,+\infty[$, the level set $E_t$
can be written as the union of a bounded set $A_t$ and a half line
$]a_t, +\infty[$. We define the rearranged set $\bar E_t = ] a_t- \vert
A_t\vert, +\infty[$ and $\bar \chi_t$ the indicator function of $\bar E_t$. Then, $\bar p(x) = \int_{0}^{+\infty} \mathrm{d}t\, \bar \chi_t(x).$

\begin{proposition}\label{lemmaRearr} Assume that the window function $w$ is symmetric decreasing\footnote{We say that a function is {\it symmetric decreasing} if it is even 
and non-increasing on the positive half line.}.
Let $f$ and $g$ be in $\mathcal{S}_f$ and $\mathcal{S}_g$ respectively, 
and let $\bar f$ and $\bar g$ be their respective increasing rearrangements. Then,
$\mathcal{W}(f,g) \geq \mathcal{W}(\bar f,\bar g).$
\end{proposition}

To prove Proposition \ref{lemmaRearr}, we make use of the 
Riesz rearrangement inequality in one dimension \cite{LiebLoss}.

Let $p$ be a non-negative measurable function on $\mathds{R}$. Then 
the \emph{symmetric decreasing rearrangement} $p^*$ of $p$ is defined as 
an even function, that is decreasing on $[0,+\infty[$ and such that the level sets  $\{x|p(x)\geq t\}$ 
and $\{x|p^*(x)\geq t\}$ have equal Lebesgue measure. Note that the decrease on $[0,+\infty[$
is not necessarily strict.

\begin{lemma}[Riesz's Inequality]\label{thmRiesz}
Let $f_1$, $f_2$, and $f_3$ be any measurable non-negative functions on the real line, and $f_1^*$, $f_2^*$, and $f_3^*$ be their \emph{symmetric decreasing} rearrangements. Then\footnote{If the left hand side is infinite so is the right hand side and the inequality is satisfied.}, 
\begin{align*}
\iint_{\mathds{R}^2} \mathrm{d}x\mathrm{d}y\, f_1(x) &f_2(x-y) f_3(y)\\
&\leq \iint_{\mathds{R}^2} \mathrm{d}x\mathrm{d}y\, f_1^*(x) f_2^*(x-y)f_3^*(y).
\end{align*}
\end{lemma}

\begin{IEEEproof}[Proof of Proposition \ref{lemmaRearr}]
Consider the expression of the potential in \eqref{eqnPotBothFts}. In order to make use of the Riesz inequality we 
 first
``symmetrize'' the profiles $f$ and $g$, and rewrite the functional in terms of symmetric profiles.
Choose $R>0$ very large but fixed. Eventually we will take $R\to +\infty$.
Denote by $\hat{f}$
the profile such that $\hat{f}(x)=f(x),\;x<R$ and $\hat{f}(x)=\hat{f}(2R-x),\;x>R$, and by $\hat{g}$
the function such that $\hat{g}(x)=g(x),\;x<R$ and $\hat{g}(x)=\hat{g}(2R-x),\;x>R$. 

We now write the potential in \eqref{eqnPotBothFts} in terms of the 
symmetrized profiles $\hat f$ and $\hat g$. We have  
\begin{align}
\mathcal{W}(f,g)&=
\lim\limits_{R\rightarrow +\infty}\int_{-\infty}^R \mathrm{d}x\Big\{I_{f,g}(x)-f(x)g^{w}(x) \Big\}
\nonumber \\
=\lim\limits_{R\rightarrow +\infty}&\bigg\{\int_{-\infty}^R \mathrm{d}x\,  I_{f,g}(x)-\int_{-\infty}^R \mathrm{d}x\,  
f(x)g^{w}(x) \bigg\}.
\label{eqnLim1}
\end{align}
For the first term in the brackets above it is straightforward to see that 
\begin{align}
\int_{-\infty}^R \mathrm{d}x\, I_{f,g}(x)=\frac{1}{2}\int_{\mathds{R}} \mathrm{d}x\, I_{\hat f,\hat g}(x).\label{eqnLim2}
\end{align}
For the second term slight care must be taken because of the convolution. We find 
\begin{align}
\int_{-\infty}^{R} \mathrm{d}x &\int_{\mathds{R}}\mathrm{d}y\, f(x)w(x-y)g(y)\label{eqnLim3}\\
&=\frac{1}{2}\int_{\mathds{R}} \mathrm{d}x\int_{\mathds{R}}\mathrm{d}y\, \hat f(x)w(x-y)\hat g(y)+o(\frac{1}{R^2}).\nonumber
\end{align}
Note that $\hat{f}$ and $\hat g$ are integrable over $\mathds{R}$. Then we can write
\begin{align*}
\mathcal{W}(f,g)&=\frac{1}{2}\lim_{R\to +\infty}(\mathcal{U}_1(\hat g) + \mathcal{U}_2(\hat f) + \mathcal{U}_3(\hat f,\hat g))
\end{align*}
where
\begin{align}
\mathcal{U}_1(\hat g)&=\int_{\mathds{R}}\mathrm{d}x\int_0^{\hat g(x)}\mathrm{d}u\, h_g^{-1}(u),\nonumber\\
\mathcal{U}_2(\hat f)&=\int_{\mathds{R}}\mathrm{d}x\int_0^{\hat f(x)}\mathrm{d}v\, h_f^{-1}(v),\nonumber\\
\mathcal{U}_3(\hat f,\hat g)&=-\int_{\mathds{R}}\mathrm{d}x\int_{\mathds{R}}\mathrm{d}y\, \hat f(x)w(x-y)\hat g(y).\label{eqnSepPot3}
\end{align}
Now consider $\hat{f}^*$ and $\hat{g}^*$ the symmetric decreasing 
rearrangements of $\hat{f}$, and $\hat{g}$, respectively. 
Each of the above terms may only decrease upon rearrangements.
Indeed each of the first two terms is a functional of a single monotone function
and thus remains unchanged by rearrangement:
$\mathcal{U}_1(\hat g)=\mathcal{U}_1(\hat g^*)$ and 
$\mathcal{U}_2(\hat f)=\mathcal{U}_2(\hat f^*)$ (see for example \cite{LiebLoss} p. 80 (3.3)). 
The term \eqref{eqnSepPot3} decreases upon rearrangement as a 
direct application\footnote{We apply this inequality for $f_1 = f$, 
$f_2 = w$, $f_3 = g$. As assumed in Proposition \ref{lemmaRearr} 
$w$ is symmetric decreasing window so that for us $w(x) = w^*(x)$ for all $x\in \mathds{R}$.} of Lemma \ref{thmRiesz}. We thus 
conclude that 
\begin{align*}
\mathcal{W}(\hat f,\hat g)\geq \frac{1}{2}\lim_{R\to +\infty}\mathcal{W}(\hat f^*, \hat g^*).
\end{align*}
To  obtain $\mathcal{W}(f,g)\geq \mathcal{W}(\bar f,\bar g)$ it remains to remark that 
$\frac{1}{2}\lim_{R\to +\infty}\mathcal{W}(\hat f^*, \hat g^*) = \mathcal{W}(\bar f,\bar g)$. This is achieved by 
reversing 
the steps \eqref{eqnLim1}-\eqref{eqnLim3}.  
\end{IEEEproof}
%

From now on we therefore restrict the functional to the spaces of non-increasing profiles.


\section{Displacement Convexity}\label{sectionDC}
A generic functional $\mathcal{F}(p)$ on a space $\mathcal{X}$ (of ``profiles" say) is said to be convex in the usual sense if for any pair $p_0,p_1\in\mathcal{X}$, and for all $\lambda\in[0,1]$, the inequality
$\mathcal{F}((1-\lambda)p_0+\lambda p_1)\leq (1-\lambda)\mathcal{F}(p_0)+\lambda \mathcal{F}(p_1)$ holds.
Displacement convexity, on the other hand, is defined as convexity under an alternative interpolation called displacement interpolation. The right setting for displacement convexity is a space of probability measures. For {\it measures over the real line} one can conveniently define the displacement interpolation in terms of the cdf's associated to the measures. This is the simplest setting and the one that we adopt here.


We think of the increasing profiles $f$ and $g$ as cdf's of some underlying measures
over the real line.
Consider two pairs $(f_0,g_0)$ and $(f_1,g_1)$, and define two (pushforward) maps $T_f$ and $T_g$ as
\begin{equation*}
T_f(x) = f_1^{-1}(f_0(x)), \,\,\, T_g(x) = g_1^{-1}(g_0(x)).
\end{equation*}
Consider the linear interpolation between points on $\mathds{R}$,
\begin{equation*}
x_{f,\lambda} = (1-\lambda)x+\lambda T_f(x), \,\,\, x_{g,\lambda} = (1-\lambda)x+\lambda T_g(x).
\end{equation*}
The displacement interpolants $(f_\lambda, g_\lambda)$ are defined so that the following equalities 
hold for all $\lambda\in [0,1]$ and $x\in \mathds{R}$
\begin{align*}
f_\lambda(x_{f,\lambda}) = f_0(x), \,\,\, g_\lambda(x_{g,\lambda}) = g_0(x).
\end{align*}
We now state the main result of this section.
\begin{proposition}\label{propConv}
The potential $\mathcal{W}(f,g)$ is displacement convex which means that for all $\lambda\in [0,1]$
\begin{align*}
\mathcal{W}(f_\lambda,g_\lambda) \leq (1-\lambda) \mathcal{W}(f_0,g_0) + \lambda \mathcal{W}(f_1,g_1).
\end{align*}
\end{proposition}

We define the two following quantities
\begin{align}
\Omega(x)=\int_{-\infty}^x\mathrm{d}z\,w(z), \qquad
V(x)=\int_{-\infty}^x\mathrm{d}z\,\Omega(z),\label{eqnDefV}
\end{align}
and call $V$ {\it the kernel} for reasons which will appear shortly.

Before proving the proposition let us first note 
{\small
\begin{align*}
&\int_{\mathds{R}}\mathrm{d}x\,(f(x)-f(+\infty))g^w(x)
=\int_{\mathds{R}}\mathrm{d}x\,(f^w(x)-f(+\infty))g(x)\\
&\; =\int_{\mathds{R}}\mathrm{d}x\int_{\mathds{R}}\mathrm{d}y\,(f(y)-f(+\infty))w(y-x)g(x)\\
&\;=
- \int_{\mathds{R}}\int_{\mathds{R}}
\mathrm{d}f(y)\,V(y-x)\,\mathrm{d}g(x),
\end{align*}}
where we have used integration by parts and $g(-\infty)=f(-\infty)=0$ for the last step.

\begin{IEEEproof}[Proof of Proposition \ref{propConv}]
Using the last identity we rewrite the potential in \eqref{eqnPotBothFts} as follows.
\begin{align}\label{eqnPotDC}
\mathcal{W}&(f,g)=\int\limits_{\mathds{R}}\mathrm{d}x\,\Bigg\{\int_0^{g(x)}\mathrm{d}u\, h_g^{-1}(u)+\int_0^{f(x)}\mathrm{d}v\, h_f^{-1}(v)\nonumber \\
&\qquad-f(+\infty)g(x)\Bigg\}+\iint_{\mathds{R}^2}
\mathrm{d}f(y)\,V(y-x)\,\mathrm{d}g(x) .
\end{align}
We now express the potential as the sum $\mathcal{W}(f,g)=\mathcal{W}_{1}(f,g)+\mathcal{W}_{2}(f,g)$, where $\mathcal{W}_{2}(f,g)$ 
consists of the last double integral in \eqref{eqnPotDC}.

We first consider $\mathcal{W}_1(f,g)$ and write
{\small
\begin{align}
&\mathcal{W}_1(f_\lambda,g_\lambda)-\mathcal{W}_1(f_0,g_0)=
\int_{\mathds{R}}\mathrm{d}x\Bigg\{\int_0^{g_\lambda(x)}\mathrm{d}u
\,h_g^{-1}(u)\label{eqn:W1}\\
&-\int_0^{g_0(x)}\mathrm{d}u
\,h_g^{-1}(u)\nonumber
+\int_0^{f_\lambda(x)}\mathrm{d}v
\,h_f^{-1}(v)
-\int_0^{f_0(x)}\mathrm{d}v
\,h_f^{-1}(v)\nonumber\\
&\qquad\qquad\qquad-\Big(f_\lambda(+\infty)g_\lambda(x)+f_0(+\infty)g_0(x) \Big)\Bigg\}.\nonumber
\end{align}
}
We remark that 
{\small
\begin{align}
\int_{\mathds{R}}&\mathrm{d}x\Bigg(\int_0^{g_\lambda(x)}\mathrm{d}u
\,h_g^{-1}(u)-\int_0^{g_0(x)}\mathrm{d}u
\,h_g^{-1}(u)\Bigg)\nonumber\\
&=\int_{\mathds{R}}\mathrm{d}x\int_0^1\mathrm{d}u\,\Big(\Theta(g_\lambda(x)-u)-\Theta(g_0(x)-u)\Big)h_g^{-1}(u),\label{eqn:W1FirstTerm}
\end{align}
}
where $\Theta$ is the Heaviside step function. 
One can check by considering the two cases $g_\lambda(x) > g_0(x)$ and $g_0(x) > g_\lambda(x)$ that
\begin{align*}
\int_0^1\mathrm{d}u\,\Big\vert\Theta(g_\lambda(x)-u)-\Theta(g_0(x)-u)\Big\vert
= \vert g_\lambda(x) - g_0(x) \vert.
\end{align*}
This observation allows us to use Fubini's theorem to swap the integrals in \eqref{eqn:W1FirstTerm}. We then write \eqref{eqn:W1FirstTerm} as
\begin{align*}
&\int_0^1  \mathrm{d}u\,h_g^{-1}(u)\int_{\mathds{R}}\mathrm{d}x\,\Big(
\Theta(g_\lambda(x)-u)-\Theta(g_0(x)-u) \Big)\\
&=\int_0^1 \mathrm{d}u\,h_g^{-1}(u)\int_{\mathds{R}}\mathrm{d}x\,\Big(
\Theta(x-g_\lambda^{-1}(u))-\Theta(x-g_0^{-1}(u)) \Big)\\
&=\int_0^1 \mathrm{d}u\,h_g^{-1}(u)(g_0^{-1}(u)-g_\lambda^{-1}(u))\\
&= \lambda\int_{\mathds{R}}\mathrm{d}g_0(x)(x-T_g(x))h_g^{-1}(g_0(x))
\end{align*}
In the last line we used a change of variables $u=g_0(x)$.
Using a similar analysis for the other terms in \eqref{eqn:W1} we find
{\small
\begin{align*}
&\mathcal{W}_1(f_\lambda,g_\lambda)-\mathcal{W}_1(f_0,g_0)\\
&=\lambda \Bigg(\int_{\mathds{R}}\mathrm{d}g_0(x)(x-T_g(x))h_g^{-1}(g_0(x))\\
&\qquad +\mathrm{d}f_0(x)(x-T_f(x))h_f^{-1}(f_0(x))+\mathrm{d}g_0(x)(x-T_g(x)) \Bigg).
\end{align*}}
We thus conclude that $\mathcal{W}_1(f_\lambda,g_\lambda)$ is linear, and hence convex, in $\lambda$.

We now consider the double integral term $\mathcal{W}_{2}(f,g)$ in \eqref{eqnPotDC}. Using again a change of variables write
\begin{align*}
\mathcal{W}&_{2}(f_\lambda,g_\lambda)=\iint_{\mathds{R}^2}
\mathrm{d}f_\lambda(y)\,V(y-x)\,\mathrm{d}g_\lambda(x)\\
&=\iint_{\mathds{R}^2}
\mathrm{d}f_0(y)\,V\Big((1-\lambda)(y-x)\nonumber\\
&\qquad\qquad\qquad\qquad+\lambda(T_f(y)-T_g(x))\Big)\,\mathrm{d}g_0(x).\nonumber
\end{align*}
This is convex in $\lambda$ because the kernel $V$ is (see \eqref{eqnDefV}). 
\end{IEEEproof}

\section{Unicity of Minimizer}\label{sectionUnicity}

\newcommand{\hf}{h_f}
\newcommand{\hg}{h_g}
\newcommand{\ff}{f}
\newcommand{\fg}{g}
\newcommand{\sptfns}{\Psi_{(-\infty,+\infty)}}
\newcommand{\smthker}{w}
\newcommand{\PhiSI}{\xi_\Phi}
\newcommand{\fS}{f^{\smthker}}
\newcommand{\gS}{g^{\smthker}}
\newcommand{\altPhi}{\phi}
\newcommand{\altPhiSI}{\xi_\phi}
\newcommand{\ind}{\boldmath{1}}
\newcommand{\indicator}[1]{\ind_{\{ #1 \}}}

In this section we prove that the potential is strictly 
displacement convex under the strictly positive gap condition. 
This implies that it admits a unique
minimizer.

Under this condition and assuming that $w$ is even and regular\footnote{Regularity of $w$ means that it is strictly positive on an interval $(-W,W),$ $W \le +\infty$
and $0$ off of $[-W,W].$}, the existence of increasing fixed point solutions was established in \cite{KRU12}.
It was also shown in \cite{KRU12} that existence of such
a fixed point implies a positive gap condition\footnote{The strictly positive gap condition requires that
$A(h_f, h_g;u) > 0$ whereas the positive gap condition requires only that $A(h_f, h_g;u) \geq 0$.}.
It was shown that if $A(h_f, h_g;u) = 0$ for some $u \in ]0,1[$ then there may be an infinite family
of fixed point solutions not equivalent under translation.
The proof of unicity relies on the potential function formulation so the
regularity conditions \eqref{eqnFGlimits} and  \eqref{eqnFGlimitsB} are required.
They can be shown to be necessary 
under mild assumptions on the scalar recursion stability of the fixed points $(0,0)$ and $(1,1)$ 
and on the decay of $w.$

From the results in the preceeding section it follows that all increasing fixed points
must have the same potential and this potential is minimal.

Let $f_0,g_0$ and $f_1,g_1$ both be non-decreasing fixed points.
We claim that they must be translates of each other, i.e., 
$y-T_f(y)$ is constant $\mathrm{d}f_0$-almost everywhere (a.e.) and
$x-T_g(x)$ takes the same constant value $\mathrm{d}g_0$-a.e.
Note that one of these conditions implies the other since both pairs are fixed points.
We will show that $y-T_f(y)$ is constant $\mathrm{d}f_0$-a.e.
The method of proof is to show that if this is not the case then 
$\mathcal{W}_{2}(f_\lambda,g_\lambda)$ is strictly convex at $\lambda = 0$
which contradicts the minimality of $f_1,g_1$. 

The proof relies on results from \cite{KRU12} that relate the 
strictly positive gap condition
to the positivity of certain integrals of spatial fixed points, which  
will be shown to imply the strict convexity of $\mathcal{W}_{2}(f_\lambda,g_\lambda)$
if $y-T_f(y)$ is not constant $\mathrm{d}f_0$-a.e. 
Since we have no further need for explicit use of $f_1,g_1$ we will simplify notation and
refer to $f_0,g_0$ as $f,g$.

Let us introduce the following functional from \cite{KRU12},
\begin{align}
\altPhiSI&(\smthker; f,g;x_1,x_2) = \label{eqn:altPhidef}\\
& \int_{0}^{+\infty}\mathrm{d}x \, \smthker(x)
\int_{]x_2,x_1+x]}\mathrm{d}f(y)\, (g(x_1+)-g(y-x)) \nonumber
\\ +&
\int_{0}^{+\infty}\mathrm{d}x \, \smthker(x)
\int_{]x_1,x_2+x]}\mathrm{d}g(y)\, (f(x_2+)-f(y-x)) \nonumber
\end{align}
where the integrals are Lebesgue-Stieltjes integrals.
If $x'<x$ then $\int_{]x,x']}\mathrm{d}g(y) \, f(y)$ is defined to be
$-\int_{]x',x]}\mathrm{d}g(y)\, f(y).$
Note that $\altPhiSI$ is non-negative; this is closely related to the strictly positive gap condition.

One of the main results in \cite{KRU12} is that for a non-decreasing fixed-point
solution we have
\begin{equation}\label{eqn:spatialintegration}
\altPhiSI(\smthker; f,g;x_1,x_2) 
=
\altPhi(h_f,h_g;g(x_1+),f(x_2+)).
\end{equation}
Note that if $x_1=x_2=x$ then the right hand side is (the right continuous version of)
$I_{f,g}(x) -f(x)g(x)$.
By the strictly positive gap condition we now have 
$\altPhiSI(\smthker; f,g;x_1,x_2) > 0$ for all $x_1,x_2$ where
$f(x_1),g(x_2) \in ]0,1[$.
Let the support of  $w$ be $[-W,W]$ (we may have $W=+\infty$).
We define $A_x=(x-W,x+W)$. From \eqref{eqn:altPhidef} it is easy to see that $\mathrm{d}g (A_x) = 0$ implies
$\altPhiSI(\smthker; f,g;x,x) = 0$ so $\mathrm{d}g (A_x) > 0$ for all $x$ with $g(x)\in ]0,1[$
and, similarly $\mathrm{d}f (A_x) > 0$ for all $x$ with $f(x) \in ]0,1[.$
These conditions imply $\dot{f}^w (x)> 0$ and $\dot{g}^w (x)> 0$
for $f(x) \in ]0,1[$ and $g(x)\in ]0,1[$ respectively, where the dot denotes differentiation.

We express $\altPhiSI$ in a more useful form for our current purpose. We assume that $x_1=x_2$ since we need the result only for this case. We claim that \eqref{eqn:altPhidef} is equal to
\begin{align}
\iint \mathrm{d}f(x) \mathrm{d}g(y)\, \mathbb{1}_{\{(x-x_1)(y-x_1)\le 0\}} \Omega(-|x-y|).
\label{eqn:altphiform}
\end{align}
To see this note that
{\small
\begin{align*}
& \int_{0}^{+\infty}\mathrm{d}x\, \smthker(x)
\int_{]x_1,x_1+x]}\mathrm{d}f(y)\, (g(x_1+)-g(y-x))
\\=&
\int_{0}^{+\infty}\mathrm{d}x\, \smthker(x)
\int_{]x_1,x_1+x]}\mathrm{d}f(y)\, \int_{]y-x,x_1]} \mathrm{d}g(z)
\\=&
\int_{0}^{+\infty} \mathrm{d}x\, \smthker(x)
\iint \mathrm{d}g(z)\mathrm{d}f(y)\,
\mathbb{1}_{\{0\le y-z \le x\}} \mathbb{1}_{\{(z-x_1)(y-x_1)\le 0\}}
\\=&
\iint \mathrm{d}g(z)\mathrm{d}f(y)\,
\mathbb{1}_{\{0\le y-z \}}\Omega(-(y-z)) \mathbb{1}_{\{(z-x_1)(y-x_1)\le 0\}}
\end{align*}
}
where for the last step we integrate by parts, using $\frac{d}{d x}( -\Omega(-x)) = w(x).$
The other term can be handled similarly and we obtain
\eqref{eqn:altphiform}.

Let $A_g$ denote the support of $\mathrm{d}g$
and let $A_f$ denote the support of $\mathrm{d}f.$
Given a real valued $s$ define measures
$\mathrm{d}f^+$ as $\mathrm{d}f^+(A) = \mathrm{d}f (A \cap \{x:x-T_f(x) \ge s \})$ and
$\mathrm{d}f^-$ as $\mathrm{d}f^-(A) = \mathrm{d}f (A \cap \{x:x-T_f(x) < s \}).$ 
Let
$A^+_f$ denote the support of $\mathrm{d}f^+$ and
$A^-_f$ denote the support of $\mathrm{d}f^-.$ 
Clearly $A_f =A_f^+\cup A_f^-.$

Since $V(x)$ is strictly convex on $]-W,W[,$ strict convexity of 
$\mathcal{W}_{2}(f_\lambda,g_\lambda)$ at $\lambda=0$ follows if there exists
$x_g \in A_g,$ $x^+_f \in A^+_f,$ and $x^-_f \in A^-_f,$ such that
$|x_g - x^+_f| < W$ and
$|x_g - x^-_f| < W.$

Assume now the existence of an $s$ so that $A^+_f$ and $A^-_f$ are both non-empty.
We will show that this implies strict convexity by establishing the existence of an $x_g$ as above.
Since this is a contradiction we conclude that $x-T_f(x)$ is constant $\mathrm{d}f$-a.e.,
giving the desired result.

If there exists $z \in A^-_f \cap A^+_f$ then we take $x_f^+ = x_f^- = z.$
Since the 
strictly positive gap condition
implies $\mathrm{d}g (z-W,z+W) >0$ (by \eqref{eqn:spatialintegration}) we have $A_g \cap (z-W,z+W) \neq \emptyset$
and we can find a suitable $x_g.$  
Assume now that $A^+_f\cap A^-_f = \emptyset.$
Let $z^- \in A^-_f$ and $z^+ \in A^+_f.$  We shall assume that
$z^- < z^+,$ the argument being the same if the order is reversed.
Define $x_f^- = \max \{ A^-_f \cap [z^-,z^+]\}$ and
$x_f^+ = \min \{ A^+_f \cap [x_f^-,z^+]\}.$
It follows that $]x_f^-,x_f^+[ \cap A_f = \emptyset.$
Since $A_f = A^+_f \cup A^-_f$ it follows from the stricly positive gap conditon \eqref{eqn:spatialintegration} that
$x_f^+ - x_f^- < 2W.$  Define $z =( x_f^+ + x_f^- )/2.$

Setting $x_1=x_2=z,$ it follows from the form \eqref{eqn:altphiform}, the 
strictly positive gap condition, and \eqref{eqn:spatialintegration} that
the $\mathrm{d}f(x)\mathrm{d}g(y)$-measure of at least one of the following
\begin{align*}
T_1 &=\{ (x,y): x\ge z,y\le z, x-y <W \}
\\
T_2 &= \{ (x,y): x\le z,y\ge z, y-x <W \}
\end{align*}
is strictly positive.
Let us assume $\mathrm{d}f\mathrm{d}g (T_1) >0.$
Note that $x_f^+ = \min A^+_f \cap [z,z+W]$ so it follows that
there exists $x_g \in ]z-W,z]$ such that $0 \le x_f^+-x_g <W.$
This clearly gives $x_f^- - x_g < W.$ In addition we have
$x_f^- -x_g > x_f^--z > -W$ so we obtain
$|x_f^- - x_g|<W.$ The argument assuming $\mathrm{d}f \mathrm{d}g (T_2)>0$ is similar.

\noindent{\bf Acknowledgments.} R. E. thanks Vahid Aref for many interesting discussions.

\vskip -1cm

\bibliographystyle{IEEEtran}
\bibliography{BIBfile}

\end{document}